\documentclass[pra,twocolumn,floatfix]{revtex4-1}
\usepackage{graphicx}% Include figure files
\usepackage{dcolumn}% Align table columns on decimal point
\usepackage{bm}% bold math
\usepackage[caption=false]{subfig}
\usepackage{color}
\usepackage[backref=none]{hyperref}
\usepackage{rotating}

\usepackage{amsfonts}
\usepackage{amssymb}
\usepackage{amsmath}
\usepackage{upgreek}

\graphicspath {{figures/}}

\hypersetup{
        colorlinks=true,          % false: boxed links; true: colored links
      linkcolor=blue,          % color of internal links
    citecolor=blue,        % color of links to bibliography
    urlcolor=green,           % color of external links
}
\makeatletter
\newcommand\numbereq{%
  \ifmeasuring@\else
    \refstepcounter{equation}%
  \fi
  \tag{\theequation}%
}

\begin{document}

%\preprint{APS/123-QED}

\title{Laser Cooling at Resonance}

\author{Yaakov~Yudkin}
%\email{yaakov.yudkin@gmail.com}
\author{Lev~Khaykovich}
%\email{lev.khaykovich@biu.ac.il}

\affiliation{Department of Physics, QUEST Center and \\ Institute of Nanotechnology and Advanced Materials, Bar-Ilan University, Ramat-Gan 5290002, Israel}

%\affiliation{

\date{\today}% It is always \today, today,

\begin{abstract}
We show experimentally that 3-D laser cooling of lithium atoms is achieved when the laser light is tuned exactly to resonance with the atomic transition. For a theoretical description of this surprising phenomenon we resolve to a full model which takes into account both the entire atomic structure and the laser light polarization. Here we build such a model for $^7$Li atoms cooled on the $D_{2}$-line in a $\sigma^+-\sigma^-$ laser configuration. We take all 24 Zeeman sub-levels into account and obtain good agreement with the experimental data. Moreover, by means of Monte-Carlo simulations we show that coherent processes play an important role in showing consistency between the theory and the experimental results. 
\end{abstract}

%\pacs{05.30.Jp 	Boson systems \\
%05.70.Ln 	Nonequilibrium and irreversible thermodynamics\\
%34.50.-s 	Scattering of atoms and molecules\\
%51.30.+i 	Thermodynamic properties, equations of state
%}% PACS, the Physics and Astronomy
\maketitle

%\footnote{dfa}

\section{Introduction}
The mechanism of Doppler cooling of neutral atoms by laser light was first suggested in the mid 70s~\citep{Haensch75,Wineland75}. 
A simplified theoretical analysis utilizes the fluctuation-dissipation theorem (FDT) and is based on finding the steady state between the friction force and the diffusion. 
Applied for a two-level atom it shows that a minimal temperature of $k_BT_D=\hbar\Gamma/2$ can be reached for a detuing $\delta=-\Gamma/2$, where $\Gamma$ is the natural linewidth of the excited state. 
This temperature is known as the Doppler limit~\citep{Gordon80,CT92}. 

A decade later these theoretical ideas were implemented experimentally on alkali atoms~\citep{Chu85}. 
But soon, to everyone's surprise, temperatures far bellow the Doppler limit were observed~\citep{Lett88}. 
This phenomenon was soon explained by the sub-Doppler cooling mechanism due to the degenerate ground states of a multilevel atom~\citep{Dalibard89}, which is what real atoms are. 
The overwhelming success of sub-Doppler cooling triggered an enormous activity in the field in the following decades paving the way for several major new research directions~\citep{CT11}.

Nevertheless, it is Doppler cooling that remains a work-horse of all laser cooling experiments with atoms and, more recently, with more and more complex molecules~\citep{Shuman10,Hummon13,Truppe17}.
This is because sub-Doppler cooling has a significantly lower velocity capture range than the Doppler cooling and it is less universal, being dependent on the ground state level degeneracy.
For example, alkaline-earth and rare-earth atoms have a non-degenerate ground state and can be expected to satisfy the conditions for Doppler cooling without a sub-Doppler mechanism setting in.
Indeed, general agreement with Doppler cooling theory has been shown in experiments with bosonic isotopes of Hg atoms~\citep{McFerran10}. 
However, in this experiment as well as in other experiments with these types of atoms, observation of the Doppler limit remained elusive~\citep{Xu02,Binnewies01,Kuwamoto99,Katori99}.
Open shell lanthanide atoms, which were introduced to laser cooling more recently, surprisingly can satisfy simple two-level atom conditions~\citep{Dreon17,Berglund08,Lu11,Aikawa12}.
Doppler theory applies well there, but again the Doppler limit was not achieved.

Thus, during the fast evolution of laser cooling, it is its most basic mechanism that has been waiting a long time for its theory to be confirmed.
Only recently it has been noticed that the special properties of metastable $^4$He atoms allow a nearly ideal realization of the Doppler cooling regime.
Thus, confirmation of several predictions of the Doppler theory in 3D were reported~\citep{Chang14}.
Lithium atoms share some of these special properties with helium atoms.
Indeed, it is well known that the sub-Doppler cooling mechanism fails in lithium as it fails in metastable helium~\footnote{We consider here the standard cooling configuration performed on the $D_2$-line. Sub-Doppler cooling of lithium atoms was demonstrated on the $D_1$-line in the so-called grey molasses regime~\citep{Grier13}. Another method to laser cool lithium atoms to lower temperatures is to use a narrow-linewidth transition for which the Doppler limit itself is lower~\citep{Duarte11}.}.
However, in contrast to helium, it is also well known that experimental results in lithium show only qualitative agreement with the Doppler theory~\citep{Schuenemann98,Mewes99}. 

Here we report one of the most intriguing and surprising deviations from the simple Doppler cooling theory in 3-D cooling of $^7$Li atoms: we observe a steady state temperature when the laser is tuned exactly to resonance with the atomic transition.
We then develop a realistic multi-level theory in a 1-D laser configuration and show good agreement with experimental results under "normal" laser cooling conditions.
In addition, we construct Monte-Carlo simulations for a better understanding of the role of coherent processes in the original theory and for simulation of 3-D cooling conditions. 
To the best of our knowledge, a realistic and quantitative theory for laser cooling of lithium has never been presented before.

Laser cooling at resonance, apart from being an interesting subject by its own rights, can be of potential interest in applications where the combination of cooling and high photon scattering rate is required.
This is the case in a recently developed advanced technique for accurate atom counting at the level of single atom resolution~\citep{Serwane11,Hume13}.
Evidently, cooling atoms at resonance, where scattering of photons is maximal, provides the best conditions for atom number counting with MOT beams.

In this paper we first describe the experiment (sec.~\ref{sec:experiment}) and demonstrate the accuracy with which the laser frequency is determined. 
In sec.~\ref{sec:theory} a realistic multi-level model is developed. 
In sec.~\ref{sec:compare} we compare experiment to theory and discuss the region of agreement and deviations from it, and in sec.~\ref{sec:MC} we describe Monte-Carlo simulations of 3-D cooling conditions. 

\section{Experiment}
\label{sec:experiment}

\subsection{Locking of the MOT Lasers}
In the experiment we collect and cool $^7$Li atoms in a magneto-optical trap (MOT) in the apparatus described elsewhere~\citep{Gross08}.
A significant improvement in the laser locking scheme has been introduced to allow precise determination of the detuning $\delta=\omega_L-\omega_A$, where $\omega_L$ is the laser frequency and $\omega_A$ is the resonance frequency of the atom. 
The previous locking scheme involved a feedback loop wired to the piezoelectric element on which the grating of the external cavity semiconductor laser is mounted. 
This method is limited to a locking bandwidth of $\sim 2$~kHz by the mechanical response of the piezo. 
Applying this technique we had obtained a laser width of $\sim 300$~kHz. 
%(Compare to the natural linewidth $\Gamma=5.9MHz$.) 
This is not good enough for investigation of the expected sharp heating of atoms in the vicinity of resonance. 
To push the laser width below $100$~kHz we wire the feedback loop to the current that runs the lasing diode. 
This way there is no (slow) mechanical response involved and a larger locking bandwidth is achieved. 
After implementing this in our system the locking bandwidth was boosted by more than a factor of 20 ($\sim 50$~kHz) and the laser linewidth was pushed down to $\sim 60$~kHz. 
This provides us with an accuracy of laser frequency determination of $0.01\;\Gamma$, where $\Gamma/2\pi=5.9$~MHz is the natural linewidth of the $D_2$-line of Li atoms.

\subsection{Locating the Resonance}

\begin{figure}
\centering\includegraphics[width=1.\columnwidth]{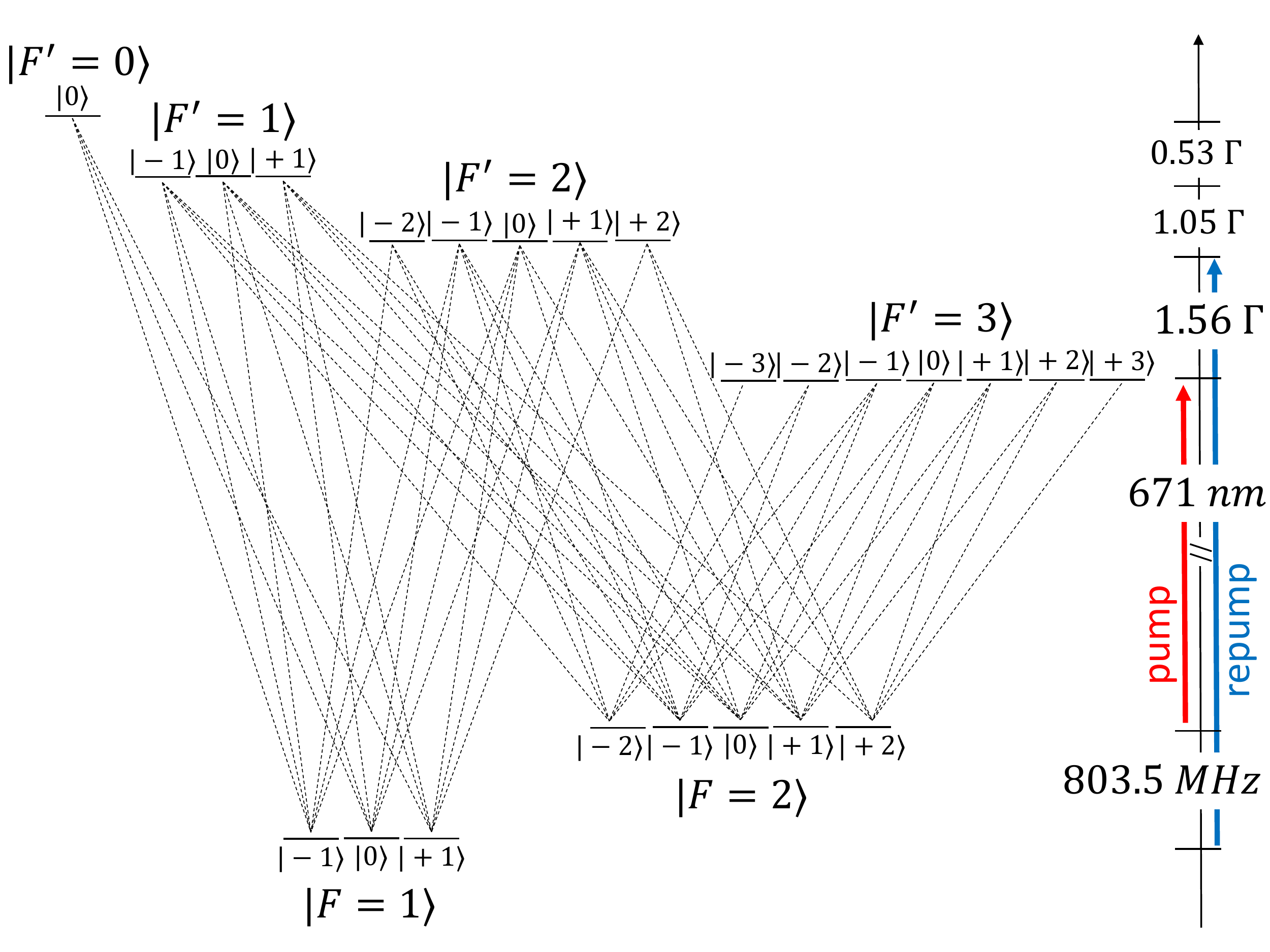}
\caption{\label{fig:D2Line} (Color online) The $D_2$-line of $^7$Li. The ground state has two hyperfine energy levels and hence both a pump (red arrow) and a repump (blue arrow) laser are needed. The excited state has four hyperfine levels. Note that their order is inverted, i.e. the lowest energy state is $|F^\prime=3\rangle$. Their energy difference is on the order of the natural linewidth $\Gamma$ making them all strongly overlapping. Each hyperfine state has degenerate Zeeman sub-levels. All allowed dipole transitions are depicted as dotted lines.}
\end{figure}

\begin{figure}
\centering\includegraphics[width=1.\columnwidth]{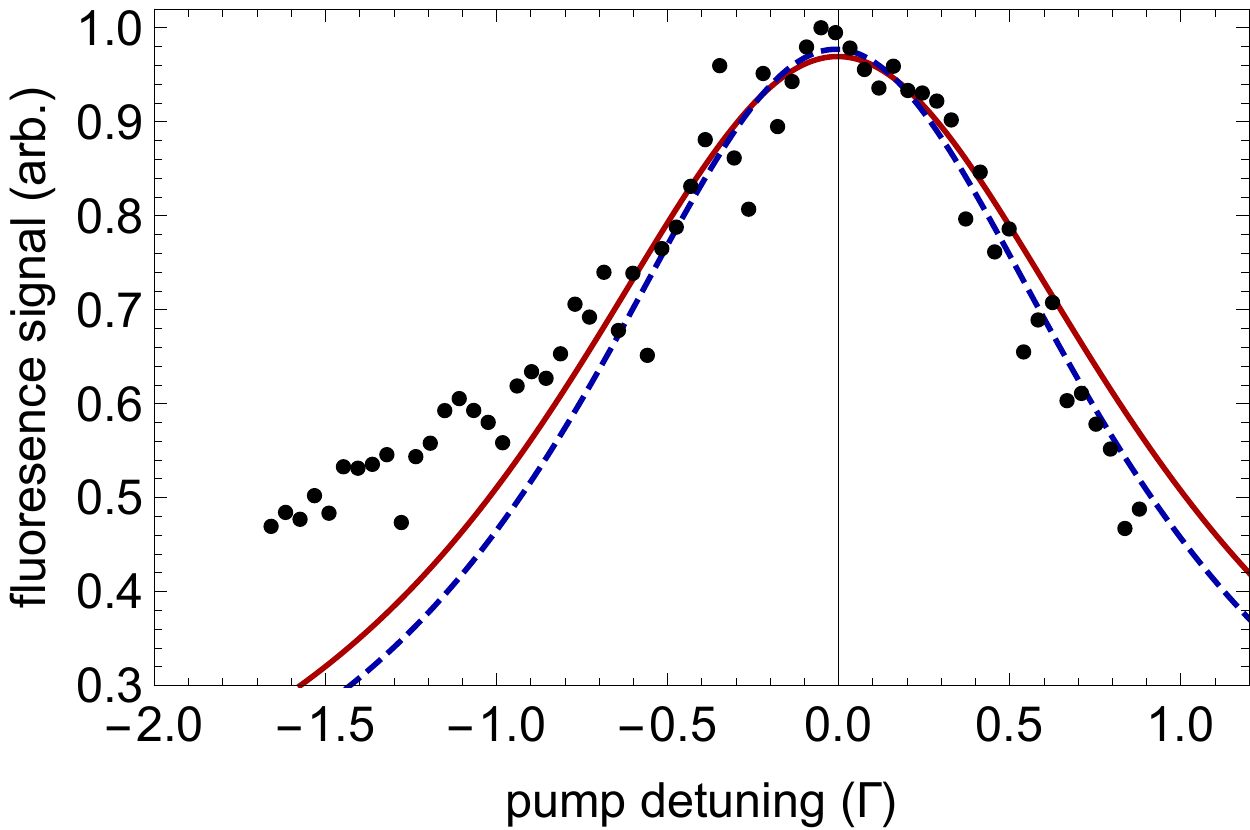}
\caption{\label{fig:resonanceCurve} (Color online) Experimentally locating the position of the resonance. Since the data is not symmetric it cannot be fitted to a Lorentzian on the entire range (see text for explanation). We use only the central points and find a very low dependence of the position of the resonance on the range chosen. The red solid line is the fit to 21 central points while the blue dashed line uses 31 points. The shift in position of the fit maximum defines the uncertainty in resonance position.}
\end{figure}

After having obtained an improved laser frequency stability we turn to precise determination of the absolute frequency. 
For this purpose, the fluorescence signal of atoms loaded into the MOT is measured as a function of the detection beam detuning. 
After being released from the optical molasses and followed by a short time of flight (TOF), the atom cloud is illuminated with a short and intense pulse of the detection light provided by the MOT pump laser only whose frequency is marked as a red arrow in Fig.~\ref{fig:D2Line}.
The pump laser intensity of the pulse per beam is $I/6=0.7\; I_{sat}$, where $I_{sat}=2.54$~mW/cm$^2$ is the saturation intensity and the pulse duration is 100~$\mu$s. 
Applying only the pump laser allows us to isolate the $|F=2\rangle \rightarrow |F^\prime=3\rangle$ closed transition from other adjacent blue detuned excited states (see Fig.~\ref{fig:D2Line}) which are open because they induce efficient optical pumping to the $|F=1\rangle$ state. 
The resulting fluorescence light is imaged onto a pco.pixelfly camera. 
The fluorescence signal as a function of the detection pulse detuning is shown in Fig.~\ref{fig:resonanceCurve}, where we define the position of the maximum as $\delta = 0$.
A perfectly isolated transition is expected to follow a symmetric Lorentzian curve.
However, the experimental data is not symmetric which is related to the optical pumping already mentioned earlier.
Above of the resonance ($\delta>0$) the laser frequency approaches the resonance transition frequency of the next hyperfine level of the excited state $|F^\prime=2\rangle$ (see fig.\ref{fig:D2Line}).
From there the atoms can decay into the lower hyperfine ground state $|F=1\rangle$. 
Since the picture is taken using the pump laser only they are now in a dark state and no longer contribute to the fluorescence signal.
Below the resonance ($\delta<0$) there are no additional energy levels so the atoms scatter a larger number of photons before an occasional optical pumping event occurs.
Therefore, the mean number of photons scattered by the atoms during the detection pulse depends on the sign of $\delta$.

As mentioned, the position of the maximum is defined as $\delta = 0$ with respect to the $|F=2\rangle\rightarrow|F^\prime=3\rangle$ transition (see Fig.\ref{fig:D2Line}) and is used as our reference point for the detuning.
In order to find it the data is fitted to a Lorentzian curve while including only the central data points and imposing zero overall offset.
The range of points used is varied and we see only small fluctuations (standard deviation of $0.02\;\Gamma$) in the position of the resonance as a function of the number of points used.
In addition, the error in the fit is typically below $0.03\;\Gamma$ so we estimate our statistical error in the determination of the resonance to be $\sim0.05\;\Gamma$.

%In addition, the residual AC Stark shift induced by the pump laser on $|F=2\rangle\rightarrow$ and $|F^\prime=3\rangle$ states causes systematic error in resonance's position determination.
Such precise determination of the resonance position requires careful considerations of possible systematic errors.
Since we calibrate the position of the resonance with a higher laser intensity compared to the optical molasses conditions (see next section), we consider the residual AC Stark shift at resonance caused by two main effects.  
The first one is related to the final velocity distribution of the atoms in the could and the second one is related to the shift of the $|F=2\rangle$ state due to the nearby allowed $|F=2\rangle\rightarrow|F^\prime=2\rangle$ transition at a detuning of $\delta=-1.5\;\Gamma$.
Both effects are estimated by calculating the expectation value of $V_{AL}^p$ (see Eq.~(\ref{eq:V_AL})) for all Zeeman sub-levels of the $|F=2\rangle$ and $|F^\prime=3\rangle$ states.
This is done by taking respective partial traces over the interaction energy multiplied by the atomic density matrix (see Sec.~\ref{sec:OBE} for definitions).
Our most conservative estimates taken at $T=2\;T_{D}$ and above mentioned laser intensity, predicts the upper bound of the systematic shift to be $-0.1\;\Gamma$.
This systematic shift is partially compensated by the conditions of the optical molasses described in the next section. 
The upper bound of the systematic error is thus reduced to $-0.05\;\Gamma$.
In addition, residual magnetic field offsets are compensated to below 10~mG. 

\subsection{Experimental Sequence and Results}

\begin{figure}
\centering\includegraphics[width=1.\columnwidth]{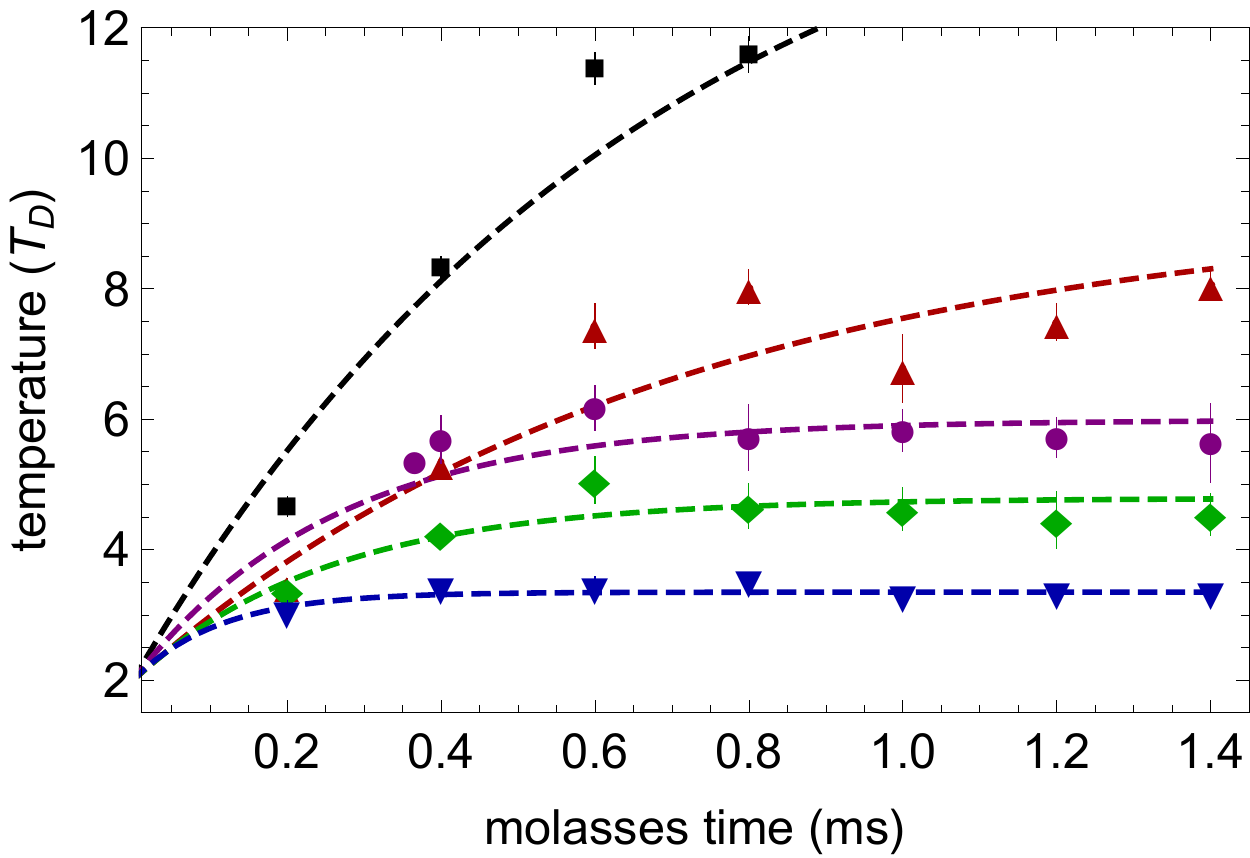}
\caption{\label{fig:TvsT} (Color online) Experimental points and fit for the temperature as a function of the duration of the molasses, after the atoms are cooled to $T_0\approx2\;T_D$ in the MOT, for different values of the pump detuning. Blue upside down triangles: $\delta/\Gamma = 0$, green diamonds: $\delta/\Gamma = 0.2$, puple circles: $\delta/\Gamma = 0.25$, red triangles: $\delta/\Gamma = 0.3$, black squares: $\delta/\Gamma = 0.34$. The steady state temperature is extracted as a fitting parameter. For $\delta=0.34\;\Gamma$ the temperature does not reach a steady state but all others do. Note that all detunings are positive.}
\end{figure}

First the MOT is loaded with $\sim 2\times 10^4$ atoms.
The detuning of the pump and repump are switched to match the conditions for minimal temperature and the atoms cooled to $\sim2T_D$ inside the MOT~\citep{Gross08}. 
This is the initial condition for the experiment. 
Then the magnetic field is turned off and the repump detuning is brought to resonance ($|F=1\rangle\rightarrow|F^\prime=2\rangle$ transition) to optimize repumping of atoms to the $|F=2\rangle$ ground state.
Simultaneously the detuning of the pump is tuned to the target value $\delta$ and the intensities of the pump and the repump are set to $I_p/6=0.06\;I_{sat}$ and $I_r/6=0.15\;I_{sat}$ respectively.
The atoms are then subject to an optical molasses for a variable time $t_{mol}$ after which a TOF measurement of the temperature is conducted.
As is clear from this description we utilize the same laser beams as for the MOT.
Therefore, the laser polarization in the molasses remains $\sigma^+ - \sigma^-$ as in the MOT. 
For each value of $\delta$ the molasses time $t_{mol}$ is varied. 
If a steady state between the force and the diffusion coefficient exists the temperature must converge towards it. 
In this way we determine the steady state temperature $T_{st}$. 
A few typical results are shown in Fig.~\ref{fig:TvsT}. 
The temperature $T(t_{mol})$ is fitted to
\begin{equation}
T(t_{mol})=T_{st}-\left(T_{st}-T_0\right)e^{-t_{mol}/\tau},
\end{equation}
where $T_{st}$ and the time constant $\tau$ are fitting parameters. 
The initial temperature $T_0=2T_D$ is set by us. 
We note that the only laser parameter being varied is thus the detuning $\delta$ of the pump, which is measured relative to the main transition of the $D_2$-line, namely $|F=2\rangle\rightarrow|F^\prime=3\rangle$. 
We have also observed, that the temperature at $\delta=0$ is largely insensitive to the repump laser detuning over a wide range of values.

\begin{figure}
\centering\includegraphics[width=1.\columnwidth]{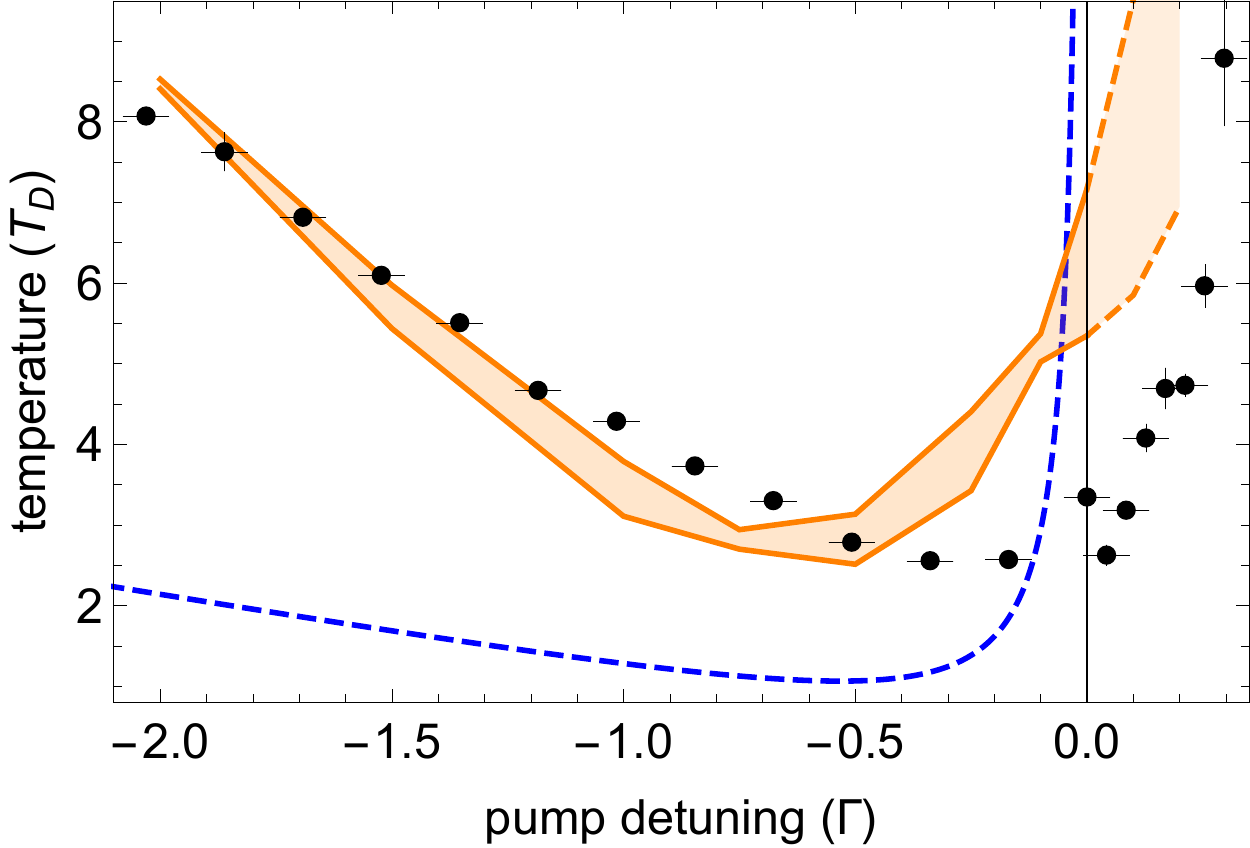}
\caption{\label{fig:TvsDET} (Color online) Experimental data points are shown as black points. The error in the detuning is $0.05\;\Gamma$ (see text). The error depicted for the temperature is one standard deviation of the fitting (dashed lines in Fig. \ref{fig:TvsT}).
The standard two-level theory is shown as a blue dashed line.
The estimated interval obtained by numerically solving the 24 level + polarization model is shown as an orange region.
The parameters for the calculation are derived from the experiment and given by $\Omega_p=0.26\;\Gamma$, $\Omega_r=0.4\;\Gamma$ and $\delta_r=0$.}
\end{figure}

In Fig.~\ref{fig:TvsDET} the fitting parameter $T_{st}$ is plotted as a function of the laser pump detuning (black points with error bars) and reflects the central result of this paper: at $\delta=0$ a steady state temperature of $\sim 3.35\;T_D$ is obtained.
This is clearly seen in Fig.~\ref{fig:TvsT} where the evolution of the temperature at $\delta=0$ is shown in blue color (upside down triangles).
Also for other blue detuned molasses, namely $\delta=0.2\;\Gamma$ (green diamonds), $\delta=0.25\;\Gamma$ (purple circles) and $\delta=0.3\;\Gamma$ (red triangles), a steady state temperature is observed.
In addition, we observe that the atomic cloud has a Gaussian distribution to a very high degree for all values of $t_{mol}$ and $\delta$ for which steady state is reached.
This is reflected by the small error bars in Fig.~\ref{fig:TvsT}.
As $\delta\rightarrow0.34\;\Gamma$ (black squares in Fig.~\ref{fig:TvsT}) no steady state is obtained and the temperature diverges (see also Fig. \ref{fig:TvsDET}).
In all of the experiments with $\delta\geq0$ the number of atoms decreases slightly (within $20\%$) as a function of the molasses time $t_{mol}$.
This loss can be attributed to an intensity imbalance of the molasses laser beams to which the system naturally becomes more sensitive at these extreme conditions. 

This deviation from a simple two-level theory of Doppler cooling is especially striking. 
We show the two-level theory prediction as a blue dashed line in Fig.~\ref{fig:TvsDET} and its complete failure to describe any of the significant features of the experimental results is educational.
We note also that the Doppler limit is unreachable experimentally. 
The minimal temperature is larger by a factor of $\sim 2$ compared to $T_D$.
It is very well known that the minimal achievable temperature increases for a large number of atoms and large densities due to photon re-scattering.
But here we work with a very small number of atoms and this effect can safely be neglected.
This factor of $~\sim 2$ is known from other experiments with $^7$Li under similar experimental conditions~\cite{Schuenemann98}.

\begin{figure}
\centering\includegraphics[width=1.\columnwidth]{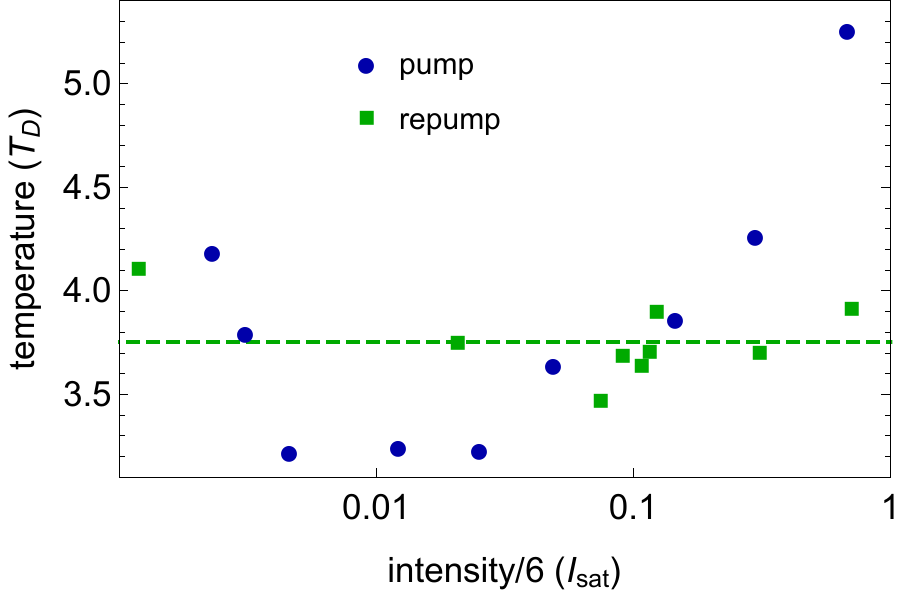}
\caption{\label{fig:TvsI} (Color online) Measurement of the steady state temperature at $\delta=0$ as a function of the pump or repump intensity. As can be seen there is an optimum in the pump intensity while the experiment is insensitive to the repump intensity. The green dashed line is the average temperature for the repump measurements (green squares).}
\end{figure}

Next, the dependence of  $T_{st}$ at resonance ($\delta=0$)  on the intensities of pump $I_p$ and repump $I_r$ is investigated.
For this, both detunings were set to resonance and the temperature was measured as a function of pump intensity $I_p$ (while $I_r/6=0.15\;I_{sat}$ as in the main experiment) and repump intensity $I_r$ (while $I_p/6=0.06\;I_{sat}$ as in the main experiment). 
The results are shown in Fig. \ref{fig:TvsI}.
Here we set the molasses time to $t_{mol}=1.4\;ms$ and find the temperature through TOF measurements.
We find that the temperature is largely insensitive to the repump intensity. 
These results indicate that the main role of the repumping laser is to prevent optical pumping to $|F=1\rangle$.
For the pump laser though, the behavior qualitatively agrees with the simple Doppler cooling model: the temperature decreases with decreasing intensity.
For too low intensities, laser cooling fails and the temperature rises again.

\section{Theory for Multi-Level Atom}
\label{sec:theory}

We describe here a 1-D semi-classical theory for laser cooling of a multi-level atom. 
As stated above, the temperature is given by the steady state of the cooling force and the diffusion.
The exact form of the force and the diffusion coefficient are found by solving the optical Bloch equations (OBE) for the relevant internal degrees of freedom at steady state~\citep{CT92}.
Then, for Doppler cooling on broad transitions ($\Gamma > \nu_r$, where $\nu_r$ is the recoil frequency) one can show that the velocity distribution function $W\left(v,t\right)$ satisfies the following Fokker-Planck (FP) equation~\citep{CT92}:
\begin{eqnarray}
\frac{\partial W(v,t)}{\partial t}&=&-\frac{\partial}{\partial (mv)}\left[F(v)W(v,t)\right]
\nonumber\\
&&+\frac{\partial^2}{\partial (mv)^2}\left[D(v)W(v,t)\right].
\label{eq:FP}
\end{eqnarray}
Its general solution in steady state ($\partial_t W(v,t)=0$) is:
\begin{equation}
W(v)=\frac{A}{D(v)}\exp\left[\int\frac{F(v)}{D(v)}d(mv)\right],
\label{eq:FP_solution}
\end{equation}
where the integration constant $A$ is used for normalization. 
The limit of vanishing velocities allows a particularly simple treatment.
To the first order in velocity, the force is linear ($F=-\alpha v$) and the diffusion coefficient is constant ($D(v)=D$) and one immediately obtains a Gaussian distribution with standard deviation $\left\langle v^2 \right\rangle=D/m\alpha$. 
The 1-D equipartition theorem (EPT) then implies the FDT:
\begin{equation}
k_BT=m\left\langle v^2 \right\rangle=\frac{D}{\alpha}.
\label{eq:steady_state_temperature}
\end{equation}
If the full velocity range is considered deviations from a Gaussian distribution become apparent for vanishing detuning even for a two-level atom~\citep{Castin89}.

In the following we first describe the OBE for lithium atoms (sec.~\ref{sec:OBE}). 
Then we find the force $F(v)$ (sec.~\ref{sec:force}) and the diffusion coefficient $D(v)$ (sec.~\ref{sec:diff}) by numerically solving the OBE. 
Then we are able to solve the FP equation resulting in the velocity distribution $W(v)$ (sec.~\ref{sec:temperature}). 
By fitting to a Gaussian we compute the width $\left\langle v^2 \right\rangle$ of the distribution and by means of Eq.~(\ref{eq:steady_state_temperature}) find the temperature. 

A few complications come up when dealing with a multi-level atom. 
For bosonic $^7$Li atoms there are a total of 24 levels (see Fig.\ref{fig:D2Line}). 
This means that we will need a $24\times24$ density matrix $\rho$ to describe the problem. 
In addition, there are two lasers in the game (pump and repump). 
The Hamiltonian will thus have two parts, one describing the pump and its interaction with the atom ($H_p$) and one for the repump ($H_r$). 
Also the laser polarization configuration has to be dealt with which leads to an extra term in the Hamiltonian.

\subsection{Optical Bloch Equations}
\label{sec:OBE}
The internal degrees of freedom of our system are described by a $24\times24$ density matrix $\rho(t)$ which satisfies the equation of motion:
\begin{equation}
\frac{d\rho}{dt}=-\frac{i}{\hbar}\left[H,\rho(t)\right]+\gamma_{dec}.
\label{eq:SE_for_DM}
\end{equation}
Here $\gamma_{dec}$ is the decay term due to spontaneous emission and the Hamiltonian is given by $H=H_p+H_r$ as indicated above. 
Each part of the Hamiltonian contains an atomic term $H_A^{p,r}$ and an interaction term $V_{AL}^{p,r}$. In the rotating frame the atomic terms are given by
\begin{equation}
H_A^{p,r}=\sum_{F^\prime,m_{F^\prime}}
\hbar\delta_{F^\prime}^{p,r}
\left|F^\prime,m_{F^\prime}\rangle\langle F^\prime,m_{F^\prime}\right|
\label{eq:H_atom}
\end{equation}
for the pump ($H_A^p$) and the repump ($H_A^r$). 
Here the sum extends over all relevant excited states ($|F^{\prime}=1,2,3\rangle$ for pump and $|F^{\prime}=0,1,2\rangle$ for repump, see Fig.~\ref{fig:D2Line}) and $\delta_{F^\prime}^p$ and $\delta_{F^\prime}^r$ are the detunings of the levels $|F^{\prime}\rangle$ for the pump and repump laser respectively. 
The interaction terms $V_{AL}^{p,r}$ are written in the long wavelength, electric dipole and rotating wave approximations. 
With help of the Rabi frequency of the pump and repump ($\Omega_p$ and $\Omega_r$) we have
\begin{eqnarray}
V_{AL}^{p,r}&=&\hbar\Omega_{p,r}\sum_{F,m_F}\sum_{F^\prime,m_{F^\prime}}
C_{\left(F^\prime,m_{F^\prime};F,m_F \right)}
\nonumber\\
&&\times\left|F^\prime,m_{F^\prime}\rangle\langle F,m_F\right|e^{ikz}+h.c.,
\label{eq:V_AL}
\end{eqnarray}
where $C_{\left(F^\prime,m_{F^\prime};F,m_F \right)}$ are the Clebsch-Gordan coefficients.
Here the sum extends over all dipole transitions (see dotted lines in Fig.\ref{fig:D2Line}).

This is the full Hamiltonian in the lab frame. 
Now we must pass into a spatially rotating frame. 
This is not to be confused with the usual (temporal) rotating frame. 
This spatial rotating frame is introduced by the laser polarization configuration. 
It is well known that a $\sigma^+ - \sigma^-$  configuration (along the $z$-axis) results in a locally linear but spatially rotating polarization. 
As the atom moves in the laser beam the direction of the linear polarization changes as a function of its location $z=vt$. 
This is best dealt with by transforming to a reference frame that rotates together with the polarization. 
Since this frame is not inertial the Hamiltonian picks up an extra term:
\begin{equation}
V_{rot}=kv\hat{J}_z,
\label{eq:V_rot}
\end{equation}
where $k=\omega/c$ is the wave number of the laser and $\hat{J}_z$ is the angular momentum operator in the $z$ direction and the generator of rotation around the $z$-axis~\citep{Dalibard89}. 
Note that this term has an explicit velocity dependence. 
The transform of the other terms of the Hamiltonian to the spatially rotating frame is trivial and will not be further discussed here. 
We just mention that it gets rid of the phase $\exp(ikz)$ in Eq.~(\ref{eq:V_AL}). 

Now the OBE at steady state can be readily obtained by plugging all Hamiltonian parts (Eqs.~(\ref{eq:H_atom}-\ref{eq:V_rot})) into Eq.~(\ref{eq:SE_for_DM}) and setting $\partial_t\rho=0$.

\subsection{Force}
\label{sec:force}

\begin{figure}
\centering\includegraphics[width=1.\columnwidth]{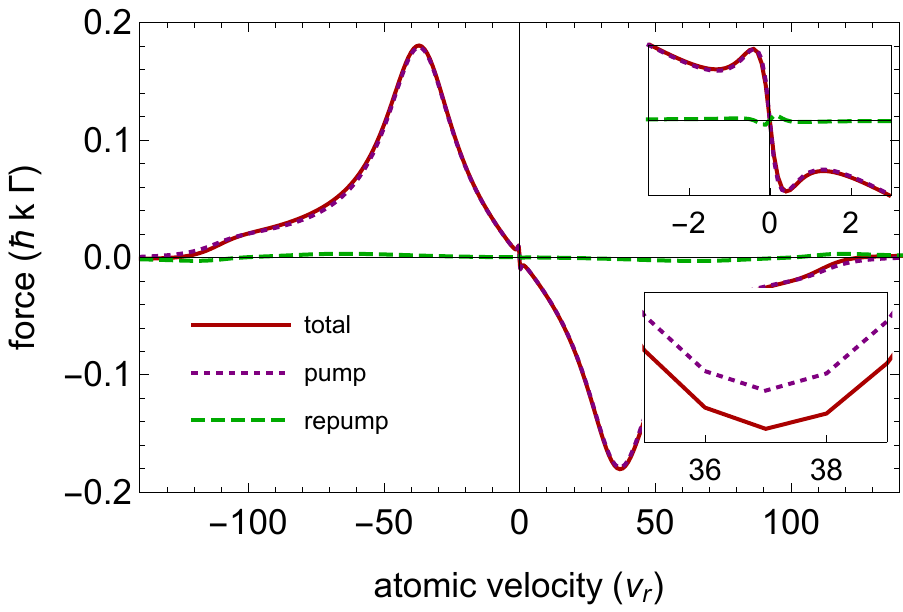}
\caption{\label{fig:forceCompare} (Color online) The numerically calculated force on a lithium atom (red, solid) for $\delta=-1.5\;\Gamma$ as a function of the velocity in units of recoil velocity $v_r=\hbar k/m$.
The Rabi frequencies $\Omega_p=0.26\;\Gamma$ and $\Omega_r=0.4\;\Gamma$ are deduced from the experiment.
The upper inset shows the sub-Doppler features of the force which are present for very low velocities.
The purple dotted line and the green dashed line are, respectively, the contributions of the pump and the repump to the total force.}
\end{figure}

\begin{figure}
\centering\includegraphics[width=1.\columnwidth]{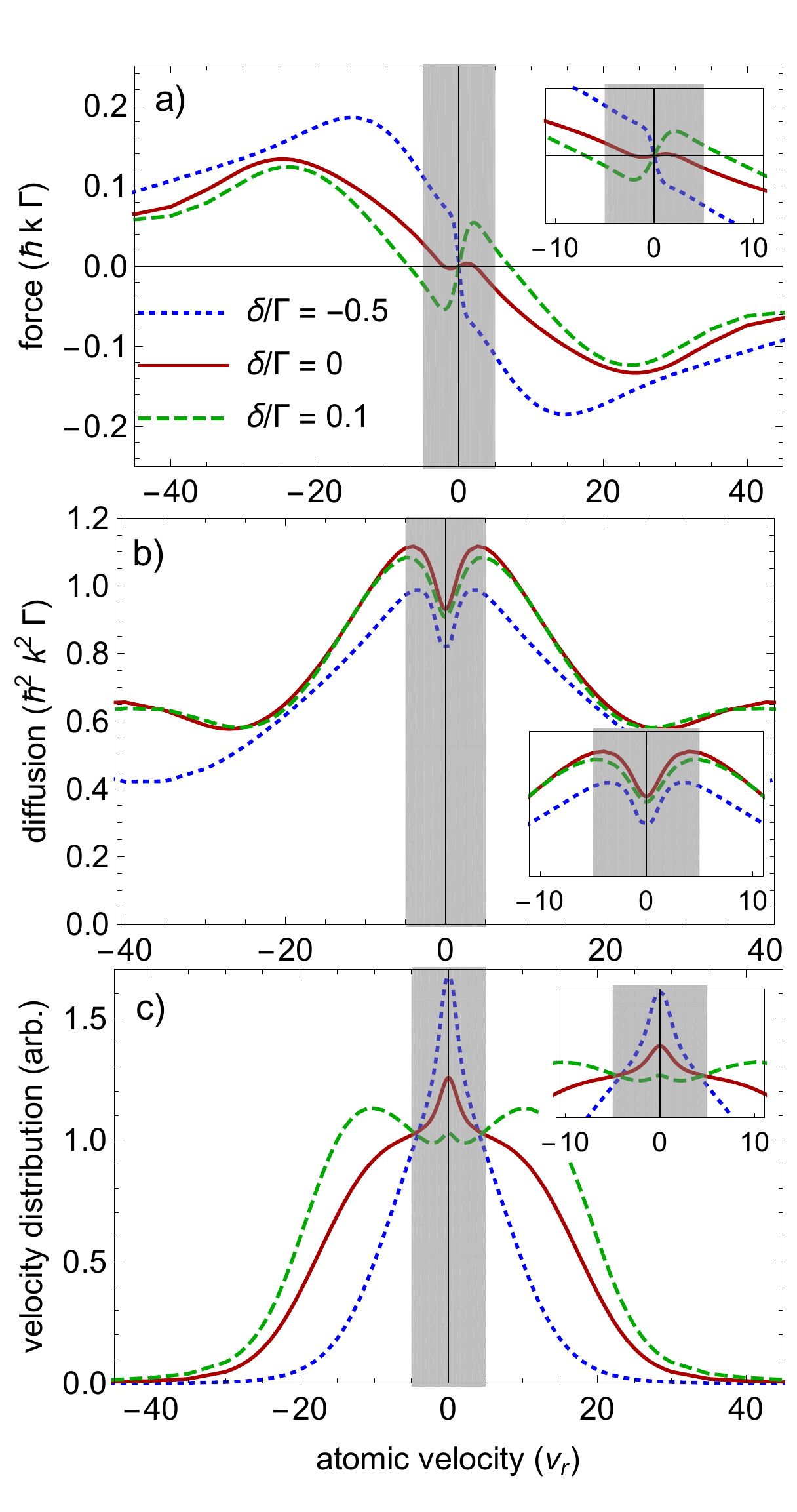}
\caption{\label{fig:force_diffusion_distribution} (Color online) Results of the numerical calculations for different values of the detuning $\delta$.
The Rabi frequencies are chosen as in Fig.~\ref{fig:forceCompare}.
The grey shaded region is $\pm5v_r$ which is equivalent to $|v|<0.7v_D$, where $v_D=\sqrt{\hbar\Gamma/m}$ is the Doppler temperature. Plot of
(a) the semi-classical force as a function of the atomic velocity, (b) the diffusion coefficient and (c) the velocity distribution for three different detunings: $\delta = -0.5\;\Gamma$ (dotted blue), $\delta=0$ (solid red) and $\delta = 0.1\;\Gamma$ (dashed green).}
\end{figure}

The force operator is given by the spatial derivative of the Hamiltonian $\hat{F}=-\nabla H$. Since we are dealing with plane waves the spatial dependence is found only in the phase $\exp(ikz)$ of Eq.~(\ref{eq:V_AL}) and the force operator is:
\begin{eqnarray}
\hat{F}&=&i\hbar k\Omega_p\sum_{pump}
C_{\left(F^\prime,m_{F^\prime};F,m_F \right)}
\left|F^\prime,m_{F^\prime}\rangle\langle F,m_F\right|
\nonumber\\
&&+i\hbar k\Omega_r\sum_{repump}
C_{\left(F^\prime,m_{F^\prime};F,m_F \right)}
\left|F^\prime,m_{F^\prime}\rangle\langle F,m_F\right|
\nonumber\\
&&+ h.c.,
\label{eq:force_operator}
\end{eqnarray}
where we have already made the transition into the spatially rotating frame. 
The semi-classical force is then obtained by taking the trace of $\hat{F}$ with the steady state solution of the OBE $\rho^{st}$:
\begin{equation}
\left\langle F \right\rangle=Tr\left(\hat{F}\rho^{st} \right).
\label{eq:force_semi_classical}
\end{equation}
This operation will single out the optical coherences of the density matrix. 
Due to the minus sign in the $h.c.$ of Eq.~(\ref{eq:force_operator}) and the overall $i$, just the imaginary part will be selected.

The numerical results are shown in Fig.~\ref{fig:forceCompare} for $\delta=-1.5\;\Gamma$.
While the red solid line shows the total force acting on an atom the purple dotted and green dashed lines show the contributions of the pump and repump laser respectively.
The sub-Doppler feature typical for a multi-level atom in a $\sigma^+ - \sigma^-$ laser configuration can clearly be seen for vanishing velocities and will be discussed shortly.
First though, notice that the force of the repump laser is negligible for all velocities. 
This was to be expected since it is tuned to resonance and explains why the experiment is insensitive to its intensity (see Fig.~\ref{fig:TvsI}).
The total force is largely dominated by the force of the pump laser.

Regarding the sub-Doppler features, we first focus on red detuned laser light (see upper inset of Fig.~\ref{fig:forceCompare} and blue dotted line in Fig.~\ref{fig:force_diffusion_distribution}(a) and its inset).
As can be seen the force strengthens as $v\rightarrow0$. 
Hence, if the force were naively expanded to first order around $v=0$ a temperature far bellow the Doppler limit would be expected. 
It is well known, however, that no sub-Doppler temperatures can be obtained for lithium atoms. 
And here the reason is visible: the region of the sub-Doppler force is extremely narrow. 
It is so narrow that the capture velocity for sub-Doppler cooling is well below the Doppler limit itself and approaches $v \sim v_r$ ($v_r=\hbar k/m$ is the recoil velocity) where the semi-classical approach becomes invalid. 
An atom at the Doppler limit has an average velocity of $v_D\approx7v_r$  and, taken in 3-D, has a negligible probability to be at the capture velocity for sub-Doppler cooling (well within the range indicated by the grey shaded region in Fig.~\ref{fig:force_diffusion_distribution}). 
Thus, the sub-Doppler mechanism is extremely inefficient in lithium and to reflect this known fact in the 1-D theory we shall cast away the shaded region from further consideration. 
This line of reasoning was first proposed in Ref~\citep{Chang14}.

When the detuning of the laser vanishes (see red line in Fig.~\ref{fig:force_diffusion_distribution}(a)) or is blue detuned (see green dashed line in Fig.~\ref{fig:force_diffusion_distribution}(a)) the slope of the force changes sign in the region $v\rightarrow0$ and therefore becomes a heating force. 
One would thus not expect a steady state temperature at all. 
But because this region is limited to within the sub-Doppler velocities we expect it to fail at heating as it fails at sub-Doppler cooling for $\delta<0$.
The negative slope found for higher velocities (Doppler region) is what the atom is subject to and therefore the force remains cooling. 
Furthermore, this slope is affected by the sub-Doppler region. 
It is steeper because it crosses zero at a finite velocity instead of zero velocity. 
In a sense, the fact that sub-Doppler cooling fails to work in lithium makes the Doppler force stronger at vanishing detuning.

The non-vanishing cooling force at resonance can be understood intuitively by looking at the level diagram of $^7$Li (Fig.\ref{fig:D2Line}). 
For $\delta<0$ all dipole transitions are red detuned and thus all contribute to a cooling force.
For $\delta = 0$ the transition $|F=2\rangle\rightarrow |F^\prime=3\rangle$ is at resonance and contributes exclusively to diffusion (heating). 
However, the transitions $|F=2\rangle\rightarrow |F^\prime=2,1\rangle$ are still red detuned. 
Hence the total force manages to maintain a damping character (negative slope). 
Because the transition $|F=2\rangle\rightarrow |F^\prime=3\rangle$ is far more dominant than the others the total force ultimately becomes a heating force for some positive detuning which we find experimentally to be $\delta \approx \Gamma/3$.

It is thus interesting to note that were the sub-Doppler mechanism efficient in lithium it would easily beat all forces originating from other excited states and then no cooling would be possible at vanishing detuning once the sub-Doppler cooling becomes heating.
We thus stress that the failure the of sub-Doppler mechanism together with the inverted order of the excited state energy levels allows the observation of finite temperatures at vanishing detuning.
This indicates the necessary conditions to observe laser cooling at resonance: small splitting and inverted character of the excited state hyperfine structure. 
%is a unique property of lithium atoms.

\subsection{Diffusion Coefficient}
\label{sec:diff}

The diffusion coefficient has two contributions. 
One comes from the coupling to the vacuum of the quantized electromagnetic field and reflects the fluctuations in spontaneous emission. 
Up to a constant it is the sum of all excited state populations
\begin{equation}
D_{vac}=\frac{1}{2}\hbar^2 k^2\Gamma\times Tr\left( \sum_{F^\prime,m_{F^\prime}}\left|F^\prime,m_{F^\prime}\rangle\langle F^\prime,m_{F^\prime}\right| \rho^{st} \right)
\label{eq:diff_vac}
\end{equation}
The second contribution is due to the fluctuations in the force. 
It is given by the integral of the two-time average~\citep{Gordon80,CT92}
\begin{eqnarray}
D_{las} &=& \Re\bigg[
\int_{0}^{\infty}d\tau
\left\langle \hat{F}(t)\hat{F}(t-\tau) \right\rangle
\nonumber\\
&&-\left\langle \hat{F}(t) \right\rangle \left\langle \hat{F}(t-\tau) \right\rangle
\bigg]
\label{eq:diff_las}
\end{eqnarray}
The key to doing this computation is the quantum regression theorem (QRT) \cite{Lax68,CT98} which states that the two-time average of two density matrix elements evolves according to the OBE, i.e. the same way as a single element.

Like the force, the diffusion coefficient $D=D_{vac}+D_{las}$ also has sub-Doppler features (Fig.~\ref{fig:force_diffusion_distribution}(b)) but, again like the force, the relevant velocities are inaccessible for the atoms. 
Also, it seems that as the velocity increases the diffusion coefficient approaches a constant value. 
This is a by product of the transformation to a spatially rotating frame which is applicable only in the low velocities limit ($kv/\Gamma \ll 1$, for $^7$Li this corresponds to $v\ll50\;v_r$). 
The actual diffusion coefficient vanishes for growing velocity.

\subsection{Extracting the Temperature}
\label{sec:temperature}

\begin{figure}
\centering\includegraphics[width=1.\columnwidth]{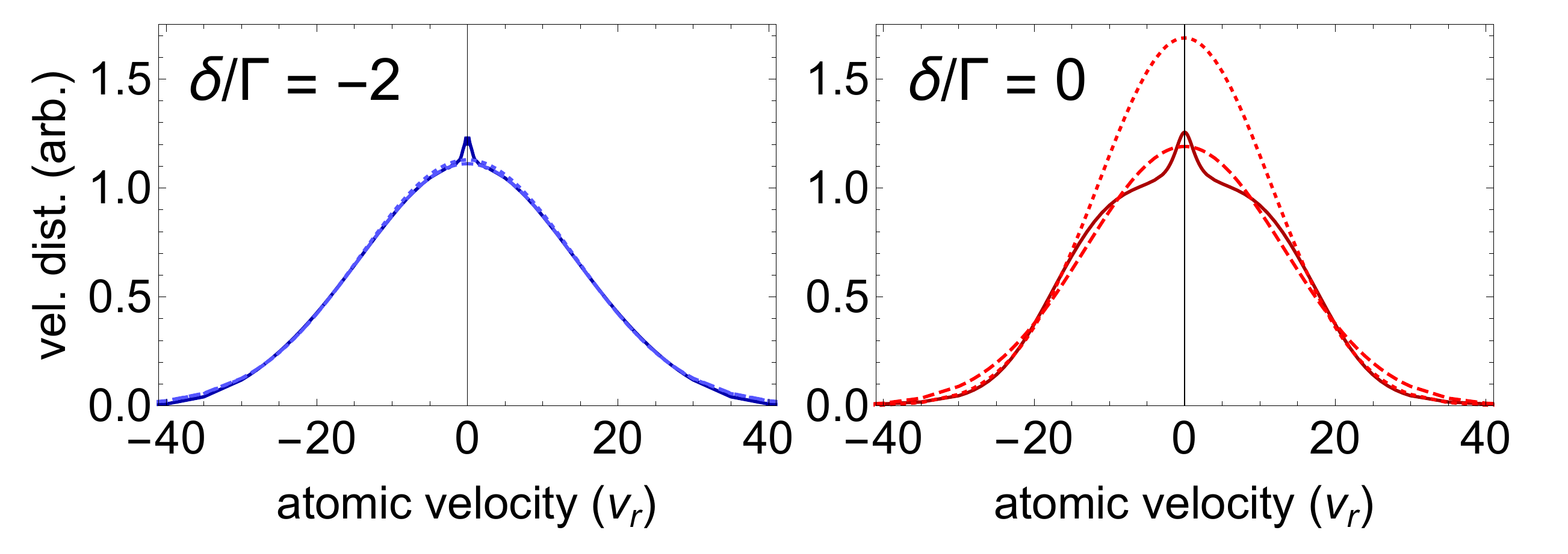}
\caption{\label{fig:fitting} (Color online) Fitting to a Gaussian of the velocity distribution. The solid line is the calculated distribution for $\delta=-2\;\Gamma$ (left) and $\delta=0$ (right). The dashed line is the fit to Gaussian excluding $\left|v\right|<5v_r$ and the dotted line for $\left|v\right|<15v_r$. This gives an estimated interval for the temperature shown in Fig.~\ref{fig:TvsDET}.}
\end{figure}

In the previous two subsections we have obtained the force $F(v)$ and the diffusion coefficient $D(v)$. 
Now we plug them into the solution of the FP equation (Eq.~(\ref{eq:FP_solution})) and solve the integral numerically. 
After normalization we find the velocity distribution $W(v)$ shown in Fig.~\ref{fig:force_diffusion_distribution}(c). 
Since only the ratio $F(v)/D(v)$ enters the exponential function in the FP equation the fact that $D(v)$ does not vanish for large velocities does not prevent convergence of $W(v)$. 
Due to the vanishing of the force, i.e. $F\left(v\rightarrow\infty\right)\rightarrow 0$, the integral will do so as well and the obtained velocity distribution behaves normally.
Although for $|v|>20v_r$ the calculated distribution might deviate from the actual distribution due to the limit of validity of the spatially rotating frame.

As can be seen the distributions are non-Gaussian. 
This is due to the sub-Doppler features in the force and the diffusion coefficient. 
Following our discussion above (Sec.~\ref{sec:force}) the sub-Doppler mechanism is inefficient in the case of lithium and the appearance of these features can be attributed to a simplified 1-D approach.
To extract meaningful information from the theory, we thus ignore the region $\left|v\right|<v_{cut}$ and fit the remaining distribution to a Gaussian. 
This is depicted in Fig.~\ref{fig:fitting} for $\delta=-2\;\Gamma$ and $\delta=0$.
Using the width $\left\langle v^2 \right\rangle$ as a fitting parameter we obtain the temperature from Eq.~(\ref{eq:steady_state_temperature}). 
We repeat this for different values of $v_{cut}$ ranging from $5\;v_r$ (dashed line in Fig.~\ref{fig:fitting}) to $1\;5v_r$ (dotted line in Fig.~\ref{fig:fitting}).

For large negative detunings ($\delta\lesssim-1.5\;\Gamma$) the Gaussian fit is very good. 
The real distribution closely resembles a Gaussian except for the regions of $v\rightarrow 0$ and $v\rightarrow\pm\infty$ which are insignificant here.
The sub-Doppler features ($v\rightarrow0$) become very narrow and small when $\delta$ is larger than the hyperfine splitting of the excited state where the multi-level atom starts resembling a three-level atom (two ground states and one excited state).
In contrast, for $\delta \gtrsim 0$ the velocity distribution deviates significantly from a Gaussian distribution. 
%There are two reasons for this. 
Here, the sub-Doppler heating force pushes the atoms away from zero velocity to a finite velocity where the force crosses zero, effectively splitting the distribution into two peaks.
Again, we attribute this effect to the limitations of an effective 1-D theory.
Also for large velocities, the diffusion coefficient even increases instead of decreasing to zero.
This is again because the spatially rotating frame approach is valid only in the limit of $v \ll 50\;v_r$. 
Due to both reasons the distribution deviates from a Gaussian and we do not expect a good correspondence between theory and experiment in this region.
However, it is interesting to note that the distribution remains finite at $\delta = 0$ (no divergence is seen) and thus the theory does predict a finite kinetic energy at resonance.

\section{Comparison of Data and Theory}
\label{sec:compare}

We calculate the force and the diffusion coefficient for pump and repump laser beam parameters which match those used in the experiment ($\Omega_p=0.26\;\Gamma$, $\Omega_r=0.4\;\Gamma$ and $\delta_r=0$ while $\delta_p=\delta$ is varied).
In Fig.~\ref{fig:TvsDET} the steady state temperature is plotted as a function of the pump laser detuning as two orange solid lines with a shaded region between them.
These two lines are derived from limiting values obtained for the range of $v_{cut}$ (see sec.~\ref{sec:temperature}).
We observe a good quantitative agreement between the experiment and the numerical calculations in the region of $\delta<-\Gamma/2$. 
This region can be trusted especially well because the sub-Doppler feature excluded from the Gaussian fit is extremely narrow and the fit quality is very good (see Fig.~\ref{fig:fitting}).
This is confirmed by the observation that the shaded region in the theory curve is very narrow here.
In addition, there is a steady state temperature all the way up to $\delta\approx+\Gamma/3$. 
The theory agrees qualitatively with this experimental fact however we do not expect quantitative agreement because of the reasons discussed in sec.~\ref{sec:temperature}.
Finally, we note that the minimum attainable temperature agrees well with the experimental value.

\section{Monte-Carlo Simulations}
\label{sec:MC}

\begin{figure}
\centering\includegraphics[width=1.\columnwidth]{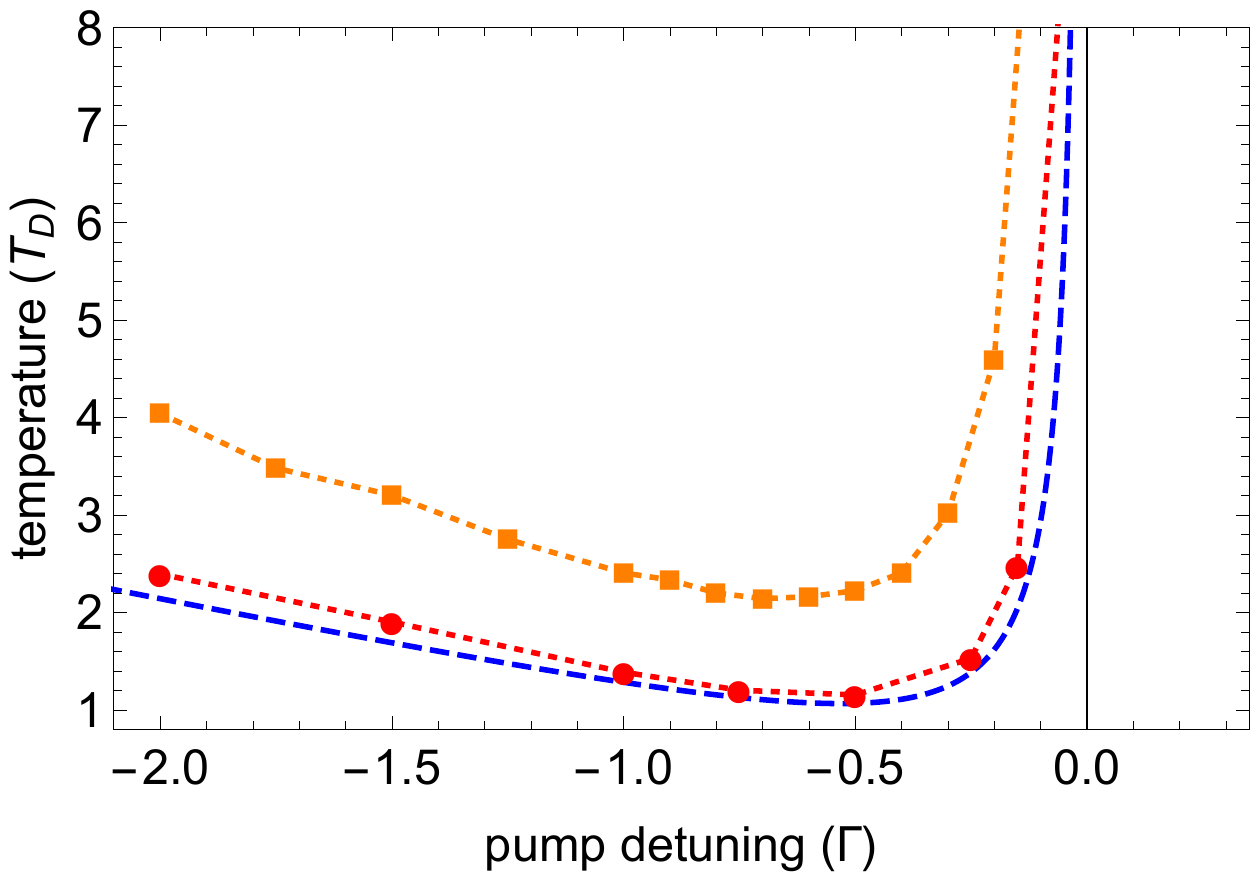}
\caption{\label{fig:MC} (Color online) Comparison of the results of a semi-classical multi-level atom Monte-Carlo simulation to the analytic solution of a two-level atom (blue, dashed).
The red dots are obtained when using random laser light polarization.
In the $\sigma^+-\sigma^-$ configuration the orange squares are obtained.
The dotted lines are to guide the eye. These results are obtained for small laser intensities so they reflect the minimal temperature.}
\end{figure}

As mentioned in the discussion of the multi-level theory, the non-Gaussian velocity distribution for all detunings shown in Fig.~\ref{fig:force_diffusion_distribution}(c) can be attributed to the theory being one dimensional.
In the experimental part (sec.~\ref{sec:experiment}) we emphasized that this non-Gaussianity is not supported by the experiments performed in 3-D.
However, extension of the semi-classical theory to a 3-D case is not within reach.
The only good alternative is to simulate the system by means of Monte-Carlo (MC) simulations.
However, building full scale 3-D MC wave function simulations which would take coherent two-photon processes responsible for sub-Doppler features in the velocity distribution  into account, is beyond the scope of the present study.
Instead, we build a semi-classical, simplified version of MC simulations (sMC) still taking into account all 24 Zeeman sublevels, the two laser frequencies and their polarizations and the 3-D character of the cooling process but including only incoherent one-photon transitions.

The purpose of the this sMC approach is twofold.
In the theoretical part of this paper (sec.~\ref{sec:theory}) we insist on artificially excluding the region of the sub-Doppler features in order to approximate the velocity distribution to a Gaussian.
It thus might be tempting to claim that coherent two-photon processes, which are responsible for these features, are not important for the treatment of the experimental results.
Therefore, the first purpose of the sMC simulations is to answer this precise question.
Then, if this question is answered positively (i.e. the coherent processes can be neglected), the second purpose is to simulate the 3-D character of the cooling.

The results of the sMC simulations are shown in Fig.~\ref{fig:MC} as orange squares and are compared to the well-known 1-D two-level theory (blue, dashed line).
Apart from a general factor of $\sim2$, both curves show identical behavior indicating that the sMC simulations miss nearly all experimental features except for one: the minimal attainable temperature is $\sim2\;T_D$ even for vanishing intensities. 
To verify the origin of this value we perform the simulations with random laser polarization.
The results are shown in Fig.~\ref{fig:MC} as red circles and rediscover the two-level system with high precision.
The extra heating present in the $\sigma^{+}-\sigma^{-}$ polarization configuration has been identified in earlier analyses of laser cooling in Refs.~\citep{Dalibard89,Castin90} as being caused by an increased step size of the random walk in momentum space.
Here we recover these results which may be considered as a candidate to explain the increased minimal temperature observed in the experiment (see Fig.~\ref{fig:TvsDET}).
However, the minimal temperature obtained in the multi-level theory (sec.~\ref{sec:theory}) taken at vanishing laser intensities does not support this claim.

Although the sMC simulations predict a finite temperature at resonance due to other hyper-fine states it exceeds the experimentally observed value by a very large factor and divergence occurs at vanishingly small positive detunings.
But even more striking is the failure of the sMC simulations to accurately predict the temperature at large and negative detunings where, naively thinking, one could consider coherent processes to play a negligible role.
Thus the sMC simulations are instructive and emphasize the role of coherent processes in the whole range of laser detunings.

\section{Conclusion}
In conclusion, we experimentally identify an unexpected regime for laser cooling of lithium atoms on the $D_2$-line: it persists up to a vanishing detuning, i.e. when the laser is tuned to resonance with the main cooling transition.
We show that a simple two-level theory is inconsistent with observations not only in this exotic regime but for all of the studied range of the laser detuning despite the known fact that sub-Doppler cooling fails to work on the $D_{2}$-line of lithium atoms.
Therefore, to describe the experiment we build a realistic theory which takes into account all 24 Zeeman sub-levels, pump and repump lasers and the laser polarization.
This theory agrees especially well with the experimental results for large and negative detunings. 
Although the theory becomes less convincing close to resonance it does predict the steady-state velocity distribution at resonance and like-wise even for small and positive detunings.
Thus, the success of cooling at resonance can be explained by the specific hyper-fine structure of the excited state of lithium's $D_{2}$-line.
On the one hand the hyper-fine structure is inverted, such that the closed transition is the lowest in energy. 
On the other, the hyper-fine splitting is very small keeping other excited states at relative proximity to the closed cooling transition.
Although this property is responsible for the failure of the sub-Doppler mechanism, it permits efficient cooling even if the laser is tuned exactly to resonance with the closed transition.

By means of sMC simulations we show that the coherent processes are crucial for explaining the experimental results. 
This is also confirmed by considering the cooling force while neglecting density matrix coherences in the multi-level theory.
In both cases we predict a finite but very large temperature of the atoms at resonance.
As a subject of future research, it is desirable to build a full MC wave function simulation in order to fully describe the experimental results. 

Cooling at resonance realizes a perfect combination of maximal photon scattering rate with effective cooling conditions.
This can be directly applied in accurate atom counting experiments with single atom resolution which would clearly benefit from this favourable combination.

Finally we note that the specific hyper-fine structure of the excited state of lithium atoms might signify that these atoms are unique to exemplify on-resonance cooling.
But since laser cooled atomic species are far from being exhausted, other such examples could be found in future research.

\section{Ackmowledgements}
We acknowledge fruitful discussions with F. Chevy, J. Dalibard and D. A. Kessler. This research was supported by the Israel Science Foundation (ISF) through grants No. 1340/16 and 2298/16. 

%\bibliography{references}% Produces the bibliography via BibTeX.
%\bibliographystyle{unsrt}

%

\end{document}